\newcommand{\CLASSINPUTbaselinestretch}{1.5}
\newcommand{\CLASSINPUTinnersidemargin}{1in}
\newcommand{\CLASSINPUTtoptextmargin}{1.65in}

\documentclass[12pt,draftclsnofoot,journal,onecolumn]{IEEEtran}
\usepackage{cite}
\ifCLASSINFOpdf
\else
\fi
\usepackage[dvips]{graphicx}
\DeclareGraphicsExtensions{.pdf,.jpeg,.png}

\usepackage[cmex10]{amsmath}
\usepackage{amssymb}
\usepackage{amsthm}

\usepackage{algorithmic}
\usepackage{array}

\usepackage{mdwmath}
\usepackage{mdwtab}
\usepackage[caption=false,font=footnotesize,subrefformat=parens,labelformat=parens]{subfig}
\usepackage{fixltx2e}
\usepackage{url}
\ifdefined\flagendfloatboulat
\usepackage[nolists,nomarkers]{endfloat}

\let\MYoriglatexcaption\caption
\renewcommand{\caption}[2][\relax]{\MYoriglatexcaption[#2]{#2}}
\fi
\usepackage{tikz-inet}
\usetikzlibrary{matrix}

\newtheorem{theorem}{Theorem}

\newtheorem*{theorem*}{Theorem}

\newtheorem{lemma}{Lemma}

\newtheorem{definition}{Definition}

\newcounter{customtheoremcounter}
\newcommand{\expected}[1]{\mathbb{E}\left[#1\right]}

\DeclareMathOperator{\sinc}{sinc}
\DeclareMathOperator{\trace}{Tr}
\DeclareMathOperator{\diag}{diag}

\DeclareMathOperator*{\argmin}{arg\,min}

\makeatletter
\def\imod#1{\allowbreak\mkern10mu({\operator@font mod}\,\,#1)}
\makeatother

\begin{document}
\title{Asymptotic Optimality of Equal Power Allocation for Linear Estimation of WSS Random Processes}

\author{Boulat~A.~Bash,~%
        Dennis~Goeckel,~%
        Don~Towsley,~%
\thanks{B. A. Bash and D. Towsley are with the Computer Science Department, University of Massachusetts, Amherst, Massachusetts.}%
\thanks{D. Goeckel is with the Electrical and Computer Engineering Department, University of Massachusetts, Amherst, Massachusetts.}%
\thanks{This research was sponsored by the National Science Foundation under 
  grants CNS-0905349 and CNS-1018464, and by the U.S. Army Research Laboratory 
  and the U.K. Ministry of Defence under Agreement Number W911NF-06-3-0001. 
  The views and conclusions contained in this document are those of the 
  author(s) and should not be interpreted as representing the official 
  policies, either expressed or implied, of the U.S. Army Research Laboratory, 
  the U.S. Government, the U.K. Ministry of Defence or the U.K.  Government. 
  The U.S. and U.K. Governments are authorized to reproduce and distribute 
  reprints for Government purposes notwithstanding any copyright notation 
  hereon.}}

\maketitle
\thispagestyle{empty}
\begin{abstract}
%
This letter establishes the asymptotic optimality of equal power allocation 
  for measurements of a continuous wide-sense stationary (WSS) random process 
  with a square-integrable autocorrelation function when linear estimation is
  used on equally-spaced measurements 
  with periodicity meeting the Nyquist criterion and with
  the variance of the noise on any sample inversely proportional to the 
  power expended by the user to obtain that measurement.

\end{abstract}
\IEEEpeerreviewmaketitle

\section{Introduction}
Linear estimation of a random process from a set of noisy measurements is 
  employed in many wireless communication and sensing applications.
We consider a scenario where a continuous 
  wide-sense stationary (WSS) random process with a square-integrable 
  autocorrelation function (ACF) is to be estimated using a Wiener filter from
  measurements that are 
  equally-spaced with periodicity meeting the Nyquist criterion and subject 
  to additive Gaussian white noise (AWGN) with variance 
  inversely proportional to the power expended by the observer to obtain it.
A Wiener filter estimator is optimal in the mean squared error
  sense if the random process is also Gaussian.
Generally, if the observation period is \emph{finite} in this setting,
  equal power allocation is suboptimal.
However, equal power allocation is simple to implement and
  performs well when the observation period is long enough to reduce
  the importance of the ``edge effects''.
In this letter we confirm this intuition by proving that the optimal power 
  allocation across equally-spaced measurements indeed tends to equality in the
  asymptotic case of an infinite number of measurements of a WSS random
  process with a square-integrable ACF.

Our scenario frequently arises in pilot symbol assisted modulation (PSAM) on 
  wireless channels.
A Gaussian random process with a square-integrable ACF
  often governs the behavior of the wireless channel,
  ensuring the optimality of Wiener filtering of the channel measurements 
  collected by the receiver from the known pilot signals inserted into 
  the transmission by the sender at the Nyquist rate for the process.
Many PSAM techniques correspond to different wireless channel models
  (see survey \cite{tong04pat} and references therein).
However, we prove a general result, confirming the intuition behind
  equally-spaced equal-power pilots used in the original work analyzing PSAM 
  \cite{cavers91pilot} (and similar approaches).

In the following section we formally state our problem.
In Section \ref{sec:prerequisites} we establish preliminary results that
  facilitate the proof of the asymptotic optimality of equal power allocation 
  for linear estimation of WSS random processes in Section \ref{sec:proof}.
Section \ref{sec:conclusion} concludes the letter.

\section{Power Allocation Problem}
\label{sec:problem}
Consider a continuous-time WSS random process $x(t)$ with a square integrable
  ACF $R_x(\tau)$:
  $\int_{-\infty}^{\infty}|R_x(\tau)|^2d\tau<\infty$.
We sample $x(t)$ at rate $\frac{1}{T_s}$ that meets the Nyquist criterion, 
  collecting $n$ samples of $x(t)$ from the interval $[0,(n-1)T_s]$.
Let $x_i^{(n)}=x(iT_s)$.
Typically (e.g.~for PSAM)
  the discrete observations of $x_i^{(n)}$ take the following form:
  $\tilde{y}_i^{(n)}=\sqrt{P_i^{(n)}}x_i^{(n)}+\tilde{z}_i^{(n)}$, where
  $\{\tilde{z}_i^{(n)}\}$ is an independent and identically distributed (i.i.d.)
  AWGN sequence with $\tilde{z}_i^{(n)}\sim \mathcal{N}(0,\sigma^2)$, and 
  $P_i^{(n)}>0$ is the power used by the observer for the $i^{\text{th}}$ 
  observation.
Normalizing the observations by $\left(P_i^{(n)}\right)^{-1/2}$ 
simplifies 
  the analysis without affecting performance,
  yielding $y_i^{(n)}=\tilde{y}_i^{(n)}\big/\sqrt{P_i^{(n)}}=x_i^{(n)}+z_i^{(n)}$,
  where $\{z_i^{(n)}\}$ is an i.i.d.~sequence with
  $z_i^{(n)}\sim\mathcal{N}(0,\sigma^2/P_i)$.
The observer is subject to the peak power constraint $P_{\max}$ such that 
  $P_i^{(n)}\leq P_{\max}$ for all $i$, where $P_{\max}$ is a finite constant.
Let $P_T(n)=\sum_{i=0}^{n-1}P_i^{(n)}$ denote the total power allocated to 
  $n$ observations of the process.

We estimate $x_i^{(n)}$ from $y_i^{(n)}$ using a Wiener filter  
  over the finite-time horizon $[0,(n-1)T_s]$ for an increasing number of 
  observations $n$.
Denote the \emph{sequence} of covariance matrices of $x_i^{(n)}$ as 
  $\{\mathbf{R}_n\}$, where $(\mathbf{R}_n)_{i,j}=R_x(|i-j|T_s)$.
Since $R_x(\tau)$ is square integrable and since  
  $x(t)$ is sampled at the Nyquist rate,
  the sequence $\{R_x(kT_s)\}_{k=0}^{n-1}$ that forms $\mathbf{R}_n$ 
  is square summable\footnote{Since $R_x(\tau)$ is square integrable and 
  band-limited to $\frac{1}{2T_s}$, it is in a Hilbert space $\mathcal{H}$. 
  A complete orthonormal basis for such is 
  $\{\sinc(\tau/T_s-k)\}_{k=-\infty}^\infty$ 
  \cite[Ch.~3.18.3]{moon00math_methods_SP}.
By Parseval's identity \cite[Th.~3.5(3)]{moon00math_methods_SP},
  $\int_{-\infty}^{\infty}|R_x(\tau)|^2d\tau=T_s^2\sum_{k=-\infty}^\infty|R_x(kT_s)|^2$.}:
  $\sum_{k=-\infty}^{\infty}|R_x(kT_s)|^2<\infty$.
Since the noise $z_i^{(n)}$ is i.i.d.,
  the sequence of covariance matrices of the observation process is 
  $\{\mathbf{R}_n+\mathbf{D}_n\}$, where 
  $\mathbf{D}_n=\diag\left(\frac{\sigma^2}{P_0^{(n)}},\frac{\sigma^2}{P_1^{(n)}},\ldots,\frac{\sigma^2}{P_{n-1}^{(n)}}\right)$
  defines a sequence of diagonal matrices.
Note that the observation process depends on the sequence of power allocation 
  vectors $\{\mathbf{p}^{(n)}\}$ through $\mathbf{D}_n$, where 
  $\mathbf{p}^{(n)}=[P_0^{(n)},P_1^{(n)},\ldots,P_{n-1}^{(n)}]$.
By  \cite[Eq.~(12.53)]{kay93statSP1},
  $\hat{\mathbf{x}}^{(n)}=\mathbf{R}_n\left(\mathbf{R}_n+\mathbf{D}_n\right)^{-1}\mathbf{y}^{(n)}$ defines the sequence of estimate vectors 
  $\{\hat{\mathbf{x}}^{(n)}\}$, where the $i^{\text{th}}$ row of matrix 
  $\mathbf{R}_n\left(\mathbf{R}_n+\mathbf{D}_n\right)^{-1}$ contains
  the Wiener filter coefficients for the estimate of $x_i^{(n)}$.
Denoting the expectation operator by $\expected{\cdot}$ and the transpose of 
  matrix $\mathbf{A}$ by $\mathbf{A}^T$, the covariance matrix
  $\mathbf{M}_n=\expected{\mathbf{e}^{(n)}\left(\mathbf{e}^{(n)}\right)^T}$ of the 
  estimate error $\mathbf{e}^{(n)}=\mathbf{x}^{(n)}-\hat{\mathbf{x}}^{(n)}$ is
  \cite[Eq.~(12.55)]{kay93statSP1}:
\begin{eqnarray}
\label{eq:M_x}\mathbf{M}_n&=&\mathbf{R}_n-\mathbf{R}_n(\mathbf{R}_n+\mathbf{D}_n)^{-1}\mathbf{R}_n\\
\label{eq:M_mtxinvlemma}&=&(\mathbf{R}_n^{-1}+\mathbf{D}_n^{-1})^{-1}
\end{eqnarray}
where (\ref{eq:M_mtxinvlemma}) is due to 
  \cite[Ch.~0.7.4]{hornjohnson85matrixanalysis}.
Again, note that the sequence $\{\mathbf{M}_n\}$ depends on the sequence of 
  power allocation vectors $\{\mathbf{p}^{(n)}\}$ via $\mathbf{D}_n$.

We are interested in the relationship between the power allocation vector
  $\mathbf{p}^{(n)}$ and the mean squared error (MSE) of 
  the estimate over the entire observation window, as the size of the 
  window, $n$, grows large.
Note that the diagonal entries of $\mathbf{M}_n$ contain the MSE of each 
  observation.
Thus, denoting the trace of the matrix $\mathbf{A}$ by $\trace[\mathbf{A}]$,
  the MSE over all observations is:
\begin{eqnarray}
\label{eq:mse_trace}\mathcal{E}\left(\mathbf{p}^{(n)}\right)\equiv\frac{1}{n}\sum_{i=0}^{n-1}\expected{e_i^2}=\frac{1}{n}\trace[\mathbf{M}_n]
\end{eqnarray}
\noindent Our main result is the following theorem:
\begin{theorem}
\label{th:main}
The MSE $\mathcal{E}\left(\mathbf{p}^{(n)}_{\text{opt}}\right)$
  of the optimal power allocation $\mathbf{p}^{(n)}_{\text{opt}}$ converges to
  the MSE $\mathcal{E}\left(\mathbf{p}^{(n)}_{\text{eq}}\right)$ of the equal 
  power allocation 
  $\mathbf{p}^{(n)}_{\text{eq}}=\{P_i^{(n)}:P_i^{(n)}=P_{\text{eq}}\}$ 
  as $n\rightarrow\infty$, with $P_{\text{eq}}=\frac{P_T(n)}{n}$.
\end{theorem}
In a typical scenario when $P_T(n)$ increases linearly with $n$, 
  $P_{\text{eq}}$ is a constant.
In order to prove this theorem in Section \ref{sec:proof}, we provide several 
  essential lemmas in the next section.

\section{Prerequisites}
\label{sec:prerequisites}
We first prove that (\ref{eq:mse_trace})
  is strictly convex over all choices of power allocation, demonstrating
  the uniqueness of the optimal power allocation.
We then introduce cyclically-symmetric functions, and prove that,
  if such functions are strictly convex on a convex domain, then the unique 
  vector that attains their minimum has equal values.
The section concludes with the introduction of the asymptotic equivalence of
  Toeplitz and circulant matrices using material from \cite{gray_toeplitz} and
  \cite{pearl73stationarydft}.

\subsection{Power Allocation Vector that Minimizes MSE is Unique}

\begin{lemma}
\label{lemma:convex}
If $\mathbf{A}$ is a real symmetric positive-definite $n\times n$ matrix, 
  then the function $f(\mathbf{x})=\trace[(\mathbf{A}+\diag(\mathbf{x}))^{-1}]$
  is strictly convex within the polytope $\sum_{i=0}^{n-1}x_i=C$, $x_i>0$.
\end{lemma}
\begin{IEEEproof}
Since the domain of $f(\mathbf{x})$ is  
  convex \cite[Ch.~2.1.2 and 2.1.4]{boyd04convexopt},
  $f(\mathbf{x})$ is strictly convex in $\mathbf{x}$ if and only if
  $g(t)=f(\mathbf{x}+t\mathbf{v})$ is strictly convex in $t$ for any
  $t\in\mathbb{R}$ and $\mathbf{v}\in\mathbb{R}^n$ such that 
  $\mathbf{x}+t\mathbf{v}$ is in the domain of $f(\mathbf{x})$ 
  (i.e. $\mathbf{x}+t\mathbf{v}$ is a real vector with positive entries that 
  sum to $C$)
  \cite[Ch.~3.1.1]{boyd04convexopt}.
This follows directly from the definition of convexity
  and is known 
  as \emph{the method of restriction to a line}.
Define $\mathbf{B}\equiv\mathbf{A}+\diag(\mathbf{x})+t\diag(\mathbf{v})$
  and note that it is symmetric positive-definite (as is $\mathbf{B}^{-1}$)
  since $\mathbf{A}$ is 
  symmetric positive-definite and $\diag(\mathbf{x})+t\diag(\mathbf{v})$ 
  is a diagonal matrix with positive entries on the diagonal.
Consider
  $h(t)=\mathbf{u}^T\mathbf{B}^{-1}\mathbf{u}$,
  where $\mathbf{u}$ is an arbitrary non-zero vector.
Its first two derivatives with respect to $t$ are 
  \cite[Ch.~D.2.1]{dattorro05convexopt}:
\begin{eqnarray}
h'(t)&=&-\mathbf{u}^T\mathbf{B}^{-1}\diag(\mathbf{v})\mathbf{B}^{-1}\mathbf{u}\\
\label{eq:d2h}h''(t)&=&2\mathbf{w}^T\mathbf{B}^{-1}\mathbf{w}
\end{eqnarray}
where $\mathbf{w}=\diag(\mathbf{v})\mathbf{B}^{-1}\mathbf{u}$.
We can substitute $\mathbf{w}$ into \eqref{eq:d2h} since $\mathbf{B}$ is 
  symmetric.
Also, since $\mathbf{B}^{-1}$ is positive-definite, $h''(t)> 0$, implying 
  that $h(t)$ is strictly convex in $t$.
Now
\begin{eqnarray}
g(t)&=&\trace[(\mathbf{A}+\diag(\mathbf{x})+t\diag(\mathbf{v}))^{-1}]\\
\label{eq:gtsum}&=&\sum_{i=0}^{n-1}\mathbf{e}_i^T(\mathbf{A}+\diag(\mathbf{x})+t\diag(\mathbf{v}))^{-1}\mathbf{e}_i
\end{eqnarray}
  where $\mathbf{e}_i$ is a vector containing one in the $i^{\text{th}}$ 
  location and zeros everywhere else.
Since each summand of (\ref{eq:gtsum}) can be written down as $h(t)$ (with 
  $\mathbf{e}_i$ replacing $\mathbf{u}$) and since the sum preserves
  convexity, $g(t)$ is strictly convex in $t$.
Therefore, $f(\mathbf{x})$ is strictly convex in $\mathbf{x}$.
\end{IEEEproof}

Since $\mathbf{R}_n$ is symmetric positive-definite,
  so is its inverse.
Also $\mathbf{D}_n^{-1}=\frac{\diag(\mathbf{p}^{(n)})}{\sigma^2}$.
Thus, by Lemma \ref{lemma:convex}, $\mathcal{E}(\mathbf{p}^{(n)})$ is strictly 
  convex and $\mathbf{p}^{(n)}_{\text{opt}}$ 
  that minimizes MSE in our problem is unique.

\subsection{Cyclically-symmetric Functions}
We next introduce a class of symmetric functions and
  prove a useful property about them.

\begin{definition}[Cyclically-symmetric function] $f(x_0,x_1,\ldots,x_{n-1})$
  is \emph{cyclically-symmetric} if 
\begin{eqnarray*}
f(x_0,x_1,\ldots,x_{n-1})=f(x_1,\ldots,x_{n-1},x_0)
\end{eqnarray*}
\end{definition}

\begin{lemma}
\label{lemma:circ}
Suppose $f(x_0,x_1,\ldots,x_{n-1})$ is strictly convex and cyclically-symmetric
  on a convex domain $\mathcal{S}$.
If vector 
  $\mathbf{x}^*=\argmin_{\mathbf{x}\in\mathcal{S}}f(x_0,x_1,\ldots,x_{n-1})$, then $x^*_0=x^*_1=\ldots=x^*_{n-1}$.
\end{lemma}
\begin{IEEEproof}
Since $f(x_0,x_1,\ldots,x_{n-1})$ is strictly convex, 
  $\mathbf{x}^*$ is unique.
Since $f(x_0,x_1,\ldots,x_{n-1})$ is cyclically-symmetric, then,
  for all $i=1,\ldots,n-1$,
  $\mathbf{x}^*$ also minimizes 
  $f(x_i,\ldots,x_{n-1},x_1,\ldots,x_{i-1})$.
Thus, $x^*_0=x^*_1=\ldots=x^*_{n-1}$.
\end{IEEEproof}

\subsection{Asymptotically Equivalent Matrices}
Results on the asymptotic equivalence of matrix sequences in 
  \cite[Ch.~2]{gray_toeplitz} enable the discussion of
  the Toeplitz and circulant matrices at the end of this section.
First, let $\mathbf{A}$ be a real-valued $n\times n$ matrix.
Then we define the matrix norms as follows:
\begin{definition}[Strong norm] 
  $\|\mathbf{A}\|=\max_{\mathbf{z}:\mathbf{z}^T\mathbf{z}=1}\left[\mathbf{z}^T\mathbf{A}^T\mathbf{A}\mathbf{z}\right]^{1/2}$.
\end{definition}
\begin{definition}[Weak norm] 
  $|\mathbf{A}|=\sqrt{\frac{1}{n}\trace\left[\mathbf{A}^H\mathbf{A}\right]}$.
\end{definition}
\noindent If $\mathbf{A}$ is symmetric positive-definite with eigenvalues 
  $\{\lambda_i\}_{i=0}^{n-1}$, 
  $|\mathbf{A}|=\sqrt{\frac{1}{n}\sum_{i=0}^{n-1}\lambda^2_i}$.
Also, $\|\mathbf{A}\|=\lambda_{\max}$ and 
  $\|\mathbf{A}^{-1}\|=1/\lambda_{\min}$,
  where $\lambda_{\max}$ and $\lambda_{\min}$ are the maximum and minimum 
  eigenvalues of $\mathbf{A}$, respectively.
\begin{lemma}[Lemma 2.3 in \cite{gray_toeplitz}]\label{lemma:mtxprodnorm}For $n\times n$ matrices $\mathbf{A}$ and $\mathbf{B}$, 
  $|\mathbf{AB}|\leq\|\mathbf{A}\|\cdot|\mathbf{B}|$.
\end{lemma}
\noindent Now define the asymptotic equivalence of matrix sequences as
  \cite[Ch.~2.3]{gray_toeplitz}:
\begin{definition}[Asymptotically Equivalent Sequences of Matrices] The
  sequences of
  $n\times n$ matrices $\{\mathbf{A}_n\}$ and $\{\mathbf{B}_n\}$ are said to be
  \emph{asymptotically equivalent} if the following hold:
\begin{eqnarray}
\label{eq:boundnorm}&\|\mathbf{A}_n\|,\|\mathbf{B}_n\|\leq M<\infty,n=1,2,\ldots&\\
&\lim_{n\rightarrow\infty}|\mathbf{A}_n-\mathbf{B}_n|=0&
\end{eqnarray}
\end{definition}
\noindent We abbreviate the asymptotic equivalence of the sequences 
  $\{\mathbf{A}_n\}$ and $\{\mathbf{B}_n\}$ by $\mathbf{A}_n\sim \mathbf{B}_n$.
Properties of asymptotic equivalence are stated and proved in 
  \cite[Theorem 2.1]{gray_toeplitz}.
A property particularly useful in the proof of Theorem \ref{th:main} is 
  re-stated here as a lemma:
\begin{lemma}
\label{lemma:aeqinv}
If $\mathbf{A}_n\sim \mathbf{B}_n$ and 
$\|\mathbf{A}_n^{-1}\|,\|\mathbf{B}_n^{-1}\|\leq K<\infty,n=1,2,\ldots$, then
$\mathbf{A}_n^{-1}\sim \mathbf{B}_n^{-1}$.
\end{lemma}
\begin{IEEEproof}
$|\mathbf{A}_n^{-1}-\mathbf{B}_n^{-1}|=|\mathbf{B}_n^{-1}\mathbf{B}_n\mathbf{A}_n^{-1}-\mathbf{B}_n^{-1}\mathbf{A}_n\mathbf{A}_n^{-1}|\leq\|\mathbf{B}_n^{-1}\|\cdot\|\mathbf{A}_n^{-1}\|\cdot|\mathbf{A}_n-\mathbf{B}_n|\xrightarrow[n\rightarrow\infty]{} 0$, 
  where the inequality is due to Lemma \ref{lemma:mtxprodnorm}.
\end{IEEEproof}
\noindent Another important consequence of asymptotic equivalence follows from
  \cite[Corollary 2.1]{gray_toeplitz}:
\begin{lemma}
\label{lemma:aeqtrace}
If $\mathbf{A}_n\sim \mathbf{B}_n$, then 
$\lim_{n\rightarrow\infty}\frac{1}{n}\trace[\mathbf{A}_n]=\lim_{n\rightarrow\infty}\frac{1}{n}\trace[\mathbf{B}_n]$ 
  when either limit exists.
\end{lemma}
\begin{IEEEproof}[Proof sketch] By the Cauchy-Schwarz inequality,
  $\left|\frac{\trace[\mathbf{A}_n-\mathbf{B}_n]}{n}\right|\leq|\mathbf{A}_n-\mathbf{B}_n|\xrightarrow[n\rightarrow\infty]{} 0$.
\end{IEEEproof}
\subsection{Sequences of Toeplitz and Circulant Matrices}
\label{sec:circ}
An $n\times n$ \emph{Toeplitz matrix} $\mathbf{T}_n$, illustrated in 
  \figurename~\subref*{fig:toeplitz}, 
  is defined by a sequence $\{t_k^{(n)}\}$ where
  $\left(\mathbf{T}_n\right)_{i,j}=t_{i-j}$.
The covariance matrix $\mathbf{R}_n$ in Section \ref{sec:problem} is Toeplitz
  and symmetric.
An $n\times n$ \emph{circulant matrix} $\mathbf{C}_n$, illustrated in 
  \figurename~\subref*{fig:circ}, 
  is defined by a sequence $\{c_k^{(n)}\}$ where 
  $\left(\mathbf{C}_n\right)_{i,j}=c^{(n)}_{(j-i)\bmod{n}}$.
Since the sequence $\{R_x(kT_s)\}_{k=0}^{n-1}$ that defines $\mathbf{R}_n$ is
  square summable, we can define  an asymptotically equivalent sequence of 
  circulant matrices $\mathbf{C}_n\sim \mathbf{R}_n$ using
  \cite[Eq.~(7)]{pearl73stationarydft}:
\begin{eqnarray}%
\label{eq:c_k}c^{(n)}_k=R_x(kT_s)+\frac{k}{n}\left(R_x((n-k)T_s)-R_x(kT_s)\right)
\end{eqnarray}
The resulting circulant matrix $\mathbf{C}_n$ is symmetric since, by 
  \eqref{eq:c_k}, $c^{(n)}_k=c^{(n)}_{n-k}$.

\begin{figure}[h]
\begin{center}
\subfloat[Toeplitz matrix]{\label{fig:toeplitz}\scalebox{0.75}{$\mathbf{T}_n=\left[\begin{array}{cccc}t_0^{(n)}&t_{-1}^{(n)}&\cdots&t_{-(n-1)}^{(n)}\\t_{1}^{(n)}&t_0^{(n)}&\cdots&t_{-(n-2)}^{(n)}\\\vdots&&\ddots&\vdots\\t_{n-1}^{(n)}&t_{n-2}^{(n)}&\cdots&t_0^{(n)}\end{array}\right]$}}
\hfil
\subfloat[Circulant matrix]{\label{fig:circ}\scalebox{0.75}{$\mathbf{C}_n=\left[\begin{array}{cccc}c_0^{(n)}&c_1^{(n)}&\cdots&c_{n-1}^{(n)}\\c_{n-1}^{(n)}&c_0^{(n)}&\cdots&c_{n-2}^{(n)}\\\vdots&&\ddots&\vdots\\c_1^{(n)}&c_{2}^{(n)}&\cdots&c_0^{(n)}\end{array}\right]$}}
\end{center}
\caption{Illustration of Toeplitz and circulant matrices.}
\end{figure}

By \cite[Eq.~(5)]{pearl73stationarydft}, 
  $\mathbf{C}_n\triangleq\mathbf{F}_n^{-1}\mathbf{\Delta}_n\mathbf{F}_n$,
  where 
  $\mathbf{\Delta}_n=\diag \left(\left\{\nu^{(n)}_i\right\}_{i=0}^{n-1}\right)$ contains 
  the diagonal entries
  $\nu^{(n)}_i=\left(\mathbf{F}_n\mathbf{R}_n\mathbf{F}_n^{-1}\right)_{i,i}$
  of the covariance matrix of the discrete Fourier
  transform (DFT) $\mathbf{F}_n\mathbf{x}^{(n)}$ of $\mathbf{x}^{(n)}$ and 
  $(\mathbf{F}_n)_{i,k}=\frac{1}{\sqrt{n}}e^{2\pi i k j/n}$ is the DFT rotation
  matrix.
Since $\mathbf{R}_n$ is positive-definite, by the properties of the similarity
  transformation, $\mathbf{F}_n\mathbf{R}_n\mathbf{F}_n^{-1}$ is 
  positive-definite and has positive diagonal entries.
Thus, $\mathbf{\Delta}_n$ is positive-definite and so is $\mathbf{C}_n$.

\section{Asymptotic Optimality of Equal Power Allocation}
\label{sec:proof}

\begin{IEEEproof}[Proof (Theorem \ref{th:main})]
Define symmetric circulant matrix $\mathbf{C}_n$ as in Section \ref{sec:circ}
  so that $\mathbf{C}_n\sim \mathbf{R}_n$ and consider 
  (\ref{eq:M_mtxinvlemma}).
Asymptotic equivalence results in the following chain of implications:
\begin{eqnarray}
\label{eq:CinvRinv}\mathbf{C}_n\sim \mathbf{R}_n&\Rightarrow&\mathbf{C}_n^{-1}\sim \mathbf{R}_n^{-1}\\
\label{eq:addD}&\Rightarrow&\mathbf{C}_n^{-1}+\mathbf{D}_n^{-1}\sim \mathbf{R}_n^{-1}+\mathbf{D}_n^{-1}\\
\label{eq:CDinvRDinv}&\Rightarrow&\left(\mathbf{C}_n^{-1}+\mathbf{D}_n^{-1}\right)^{-1}\sim \left(\mathbf{R}_n^{-1}+\mathbf{D}_n^{-1}\right)^{-1}\\
\label{eq:circMlimit}&\Rightarrow&\lim_{n\rightarrow\infty}\frac{1}{n}\trace[\mathbf{L}_n]=\lim_{n\rightarrow\infty}\frac{1}{n}\trace[\mathbf{M}_n]
\end{eqnarray}
where $\mathbf{L}_n\equiv\left(\mathbf{C}_n^{-1}+\mathbf{D}_n^{-1}\right)^{-1}$.
Since $\mathbf{R}_n$ and $\mathbf{C}_n$ are symmetric positive-definite, the 
  conditions for Lemma \ref{lemma:aeqinv} hold, resulting in
  (\ref{eq:CinvRinv}).
Then (\ref{eq:addD}) follows from adding $\mathbf{D}_n^{-1}$ to both sides of 
  the asymptotic equivalence relation
  and noting that the condition (\ref{eq:boundnorm}) is satisfied via 
  Weyl's inequality \cite[Theorem 4.3.1]{hornjohnson85matrixanalysis}
  since the peak power constraint on the observer implies that
  $\mathbf{D}_n^{-1}$ has finite eigenvalues.
Lemma \ref{lemma:aeqinv} yields (\ref{eq:CDinvRDinv}), and
  (\ref{eq:circMlimit}) is due to Lemma \ref{lemma:aeqtrace}. 
Let
\begin{eqnarray}
\label{eq:E_equiv}\mathcal{E}_{\text{equiv}}(\mathbf{p}^{(n)})&\equiv&\frac{1}{n}\trace[(\mathbf{C}_n^{-1}+\mathbf{D}_n^{-1})^{-1}]
\end{eqnarray}
\noindent Then, since we defined $\mathcal{E}(\mathbf{p}^{(n)})\equiv\frac{1}{n}\trace[\mathbf{M}_n]$ in \eqref{eq:mse_trace}, (\ref{eq:circMlimit}) can be restated as follows:
\begin{eqnarray}
\label{eq:aeqmse}
\lim_{n\rightarrow\infty}\mathcal{E}(\mathbf{p}^{(n)})&=&\lim_{n\rightarrow\infty}\mathcal{E}_{\text{equiv}}(\mathbf{p}^{(n)})
\end{eqnarray}
\noindent Since $\mathbf{C}_n$ is symmetric positive-definite,
  by Lemma \ref{lemma:convex},  
  $\mathcal{E}_{\text{equiv}}(\mathbf{p}^{(n)})$ is strictly convex in
  $\mathbf{p}^{(n)}$. 
Showing that $\mathcal{E}_{\text{equiv}}(\mathbf{p}^{(n)})$ is 
  cyclically-symmetric with respect to 
  $\mathbf{p}^{(n)}$ would complete the proof by Lemma \ref{lemma:circ}.
The discussion of the convergence of $\mathcal{E}(\mathbf{p}^{(n)})$ to
  $\mathcal{E}_{\text{equiv}}(\mathbf{p}^{(n)})$ follows the proof.

Denote the similarity transformation 
$\mathcal{R}_i(\mathbf{A})\triangleq\mathbf{S}_i\mathbf{A}\mathbf{S}_i^{-1}$
  of an $n\times n$ matrix $\mathbf{A}$ \emph{the rotation of degree $i$},
where 
$\mathbf{S}_i=\left[\begin{array}{cc}\mathbf{0}&\mathbf{I}_{(n-i)\times(n-i)}\\\mathbf{I}_{i\times i}&\mathbf{0}\end{array}\right]$
  and $\mathbf{I}_{n\times n}$ is an $n\times n$ identity matrix.
$\mathbf{S}_i$ is a permutation matrix, and is thus orthogonal,
  implying that $\mathbf{S}_i^{-1}=\mathbf{S}_i^{T}=\mathbf{S}_{n-i}$.
Suppose that the rows and columns of matrix $\mathbf{A}$ are labeled 
  $0,\ldots,n-1$ top-to-bottom and right-to-left, respectively.
Then $\mathbf{S}_i\mathbf{A}$ produces a matrix with the top $i$ rows of
  $\mathbf{A}$ shifted to the bottom (i.e.~rows $0,\ldots,i-1$ become rows 
  $n-i-1,\ldots,n-1$), and $\mathbf{A}\mathbf{S}_i^{-1}$ 
  produces a matrix with the left $i$ columns of $\mathbf{A}$ shifted to the
  right (i.e.~columns $0,\ldots,i-1$ become columns $n-i-1,\ldots,n-1$).
Shifting the top $i$ rows of a circulant matrix $\mathbf{C}$ down
  produces the same matrix as shifting the left $n-i$ columns to the right.
Thus, $\mathbf{S}_i\mathbf{C}=\mathbf{C}\mathbf{S}_i$, which implies
  the \emph{rotation invariance} of circulant matrices: 
  $\mathcal{R}_i(\mathbf{C})=\mathbf{C}$.

Inverse $\mathbf{A}^{-1}$ of matrix $\mathbf{A}$ can be expressed as
  $\left(\mathbf{A}^{-1}\right)_{i,j}=\frac{(-1)^{i+j}}{\det(\mathbf{A})}\mathcal{M}_{j,i}(\mathbf{A})$
  where $\det(\mathbf{A})$ denotes the determinant of $\mathbf{A}$ and 
  $\mathcal{M}_{i,j}(\mathbf{A})=\det(\mathbf{A}^{(i,j)})$ with 
  the sub-matrix $\mathbf{A}^{(i,j)}$ formed by removing
  row $i$ and column $j$ from $\mathbf{A}$
  \cite[Ch.~0.8.1]{hornjohnson85matrixanalysis}.
Thus, (\ref{eq:E_equiv}) can be re-stated as:
\begin{eqnarray}
\label{eq:traceL}\mathcal{E}_{\text{equiv}}(\mathbf{p}^{(n)})&=&\frac{\sum_{k=0}^{n-1}\mathcal{M}_{k,k}(\mathbf{C}_n^{-1}+\mathbf{D}_n^{-1})}{n\det(\mathbf{C}_n^{-1}+\mathbf{D}_n^{-1})}
\end{eqnarray}

The inverse of a circulant matrix, if it exists, is 
  circulant\footnote{We note that, in 
  general, the inverses of Toeplitz matrices are \emph{not} Toeplitz.}
  \cite[Theorem 3.1 (3)]{gray_toeplitz}.
Due to the rotational invariance of circulant matrices, for all 
  $i=1,\ldots,n-1$,
  $\mathcal{R}_i(\mathbf{C}^{-1}_n+\mathbf{D}^{-1}_n)=\mathbf{C}^{-1}_n+
  \mathcal{R}_i(\mathbf{D}^{-1}_n)$ with $\mathcal{R}_i(\mathbf{D}^{-1}_n)=
  \diag(\frac{P_i^{(n)}}{\sigma^2},\ldots,\frac{P_{n-1}^{(n)}}{\sigma^2},\frac{P_0^{(n)}}{\sigma^2},\ldots,\frac{P_{i-1}^{(n)}}{\sigma^2})$.
The denominator of (\ref{eq:traceL}) is cyclically-symmetric with respect
  to $\mathbf{p}^{(n)}$ due to the rotation being a similarity transformation, 
  which preserves the determinant.
Since the submatrix of 
  $\mathbf{C}^{-1}_n+\mathbf{D}^{-1}_n$ with row $k$ and column $k$
  removed is a submatrix of 
  $\mathbf{C}^{-1}_n+\mathcal{R}_i(\mathbf{D}^{-1}_n)$ with 
  row $(k-i)\bmod{n}$ and column $(k-i)\bmod{n}$ removed, 
  the numerator of (\ref{eq:traceL}) is also cyclically-symmetric with respect
  to $\mathbf{p}^{(n)}$.
Therefore, $\mathcal{E}_{\text{equiv}}(\mathbf{p}^{(n)})$ is 
  cyclically-symmetric and, by Lemma 
  \ref{lemma:circ}, $\argmin_{\mathbf{p}^{(n)}}\mathcal{E}_{\text{equiv}}(\mathbf{p}^{(n)})=\mathbf{p}^{(n)}_{\text{eq}}$.
By (\ref{eq:aeqmse}), $\lim_{n\rightarrow\infty}\mathcal{E}(\mathbf{p}^{(n)}_{\text{eq}})=\lim_{n\rightarrow\infty}\mathcal{E}_{\text{equiv}}(\mathbf{p}^{(n)}_{\text{eq}})$,
  completing the proof.
\end{IEEEproof}

\emph{Convergence:}
First bound $|\mathcal{E}(\mathbf{p}^{(n)})-
  \mathcal{E}_{\text{equiv}}(\mathbf{p}^{(n)})|
  =\left|\frac{1}{n}\trace[\mathbf{L}_n-\mathbf{M}_n]\right|
  \leq\left|\mathbf{L}_n-\mathbf{M}_n\right|$ using the
  Lemma \ref{lemma:aeqtrace} proof idea.
Now, as done in the proof of Lemma \ref{lemma:aeqinv}, apply Lemma 
\ref{lemma:mtxprodnorm}:
  $\left|\mathbf{L}_n-\mathbf{M}_n\right|\leq \|\mathbf{L}_n\|\cdot\|\mathbf{M}_n\|\cdot\|\mathbf{C}_n^{-1}\|\cdot\|\mathbf{R}_n^{-1}\|\cdot|\mathbf{C}_n-\mathbf{R}_n|$.
The strong norm terms are bounded by Weyl's theorem per the arguments following
  \eqref{eq:circMlimit}, while the discussion following  
  \cite[Eq.~(10)]{pearl73stationarydft} asserts that
  $|\mathbf{C}_n-\mathbf{R}_n|\leq\frac{C}{\sqrt{n}}+\epsilon$ where $C$ and
  $\epsilon$ are constants, with $\epsilon$ arbitrarily small.
Thus, in \eqref{eq:aeqmse}, $\mathcal{E}(\mathbf{p}^{(n)})$ and 
  $\mathcal{E}_{\text{equiv}}(\mathbf{p}^{(n)})$ converge at a
  rate proportional to $\frac{1}{\sqrt{n}}$.

Finally, while the numerical results are not in the scope of this letter, our 
  evaluations show that 
  $\mathcal{E}(\mathbf{p}^{(n)}_{\text{eq}})$ converges to 
  $\mathcal{E}(\mathbf{p}^{(n)}_{\text{opt}})$ fairly 
  quickly\footnote{E.g.: %
  We used MATLAB R2011a \texttt{fmincon} at default settings with $P_T(n)=n$ 
  and $\sigma^2=1$, initialized by $\mathbf{p}^{(n)}_{\text{eq}}$, to find 
  $\mathcal{E}(\mathbf{p}^{(n)}_{\text{opt}})$ for $R_x(\tau)=e^{-|\tau|}$, 
  where $x(t)$ is sampled at the rate
  $\frac{\tan\frac{0.99\pi}{2}}{\pi}$ Hz (capturing 99\% of the spectrum).
  This converged (i.e.~\texttt{fmincon} was not able to improve 
  $\mathcal{E}(\mathbf{p}^{(n)}_{\text{opt}})$ below the initial
  $\mathcal{E}(\mathbf{p}^{(n)}_{\text{eq}})$ within the default tolerance 
  $1\times10^{-6}$) at $n=84$.
  For Jakes' model with $R_x(\tau)=J_0(2\pi f_D \tau)$ sampled at the rate $2f_D$ Hz (Nyquist),
      convergence was immediate (at $n=1$).}
  in typical wireless communication scenarios.

\section{Conclusion}
\label{sec:conclusion}
The asymptotic optimality is established for the equal power allocation between 
  equally-spaced measurements used for Wiener filter estimation of a continuous 
  WSS random process with a square-integrable ACF, where the periodicity of 
  the measurements meets the Nyquist criterion
  and the measurements are
  subject to AWGN with variance inversely proportional to the power 
  expended by the observer.

\bibliographystyle{IEEEtran}
\bibliography{IEEEabrv,../../../papers}
\end{document}